# Superconductivity in the pressure-amorphized topological insulator $CrP_4$

Chutong Zhang[1,#], Xiangzhuo Xing[1,2,#,*], Na Zuo[1,#], Bowen Zheng[1,#], Bin Li[3,*], Jiajia Feng[4], Xiaolei Yi[5], Yan Meng[6], Xiaoran Zhang[1,2], Bingchao Yang[1,2], Chao Wang[1,2], Xin Chen[1,2], Yongsheng Zhang[1,2], Xiaofeng Xu[7,*] and Xiaobing Liu[1,2,*]

[1] *Key Laboratory of Quantum Materials under Extreme Conditions in Shandong Province, School of Physics and Physical Engineering, Qufu Normal University, Qufu 273165, China*

[2] *Laboratory of High Pressure Physics and Material Science (HPPMS), Advanced Research Institute of Multidisciplinary Sciences, Qufu Normal University, Qufu 273165, China*

[3] *School of Science, Nanjing University of Posts and Telecommunications, Nanjing 210023, China*

[4] *Center for High Pressure Science and Technology Advanced Research, Beijing 100193, China*

[5] *College of Physics and Electronic Engineering, Xinyang Normal University, Xinyang 464000, China*

[6] *School of Physics and Electronic Engineering, Jining University, Qufu 273155, China*

[7] *School of Physics, Zhejiang University of Technology, Hangzhou 310023, China*

[#] These authors contributed equally to this work.

*Corresponding authors, Email: xzxing@qfnu.edu.cn, libin@njupt.edu.cn, xuxiaofeng@zjut.edu.cn, xiaobing.phy@qfnu.edu.cn

## ABSTRACT

The interplay among superconductivity, magnetism, and nontrivial band topology represents one of the most compelling frontiers in condensed matter physics. The exploration of novel superconductivity in 3*d* transition-metal compounds, particularly the rare Cr-based systems containing strongly magnetic Cr ions, has long attracted attention owing to their unconventional pairing mechanisms that challenge conventional wisdom. Yet, Cr-based superconductors remain scarce, especially those possessing nontrivial topological character, underscoring the urgent need to uncover new members. Here we report the observation of superconductivity in pressure-amorphized Cr-based topological insulator $CrP_4$. Upon compression, $CrP_4$ undergoes an anomalous quantum phase transition from a metallic to a semiconducting-like state at around 15 GPa, driven by significant changes in the electronic structure. At approximately 70 GPa, re-metallization with superconductivity occurs alongside an irreversible amorphization. The superconducting transition temperature $T_c$ increases monotonically with pressure, reaching 4.8 K at 141.3 GPa. Furthermore, theoretical calculations predict multiple topological phase transitions from a strong topological insulator to a trivial state and finally back to a strong topological state under pressure. Our study not only establishes $CrP_4$ as the first Cr-based amorphous superconductor but also opens a new paradigm for exploring superconducting and topological properties in amorphous materials.

## I. INTRODUCTION

Transition-metal oxide and pnictide superconductors, exemplified by cuprates [1], iron pnictides [2], and more recently nickelates [3], have attracted enduring attention owing to their intertwined electronic orders and rich quantum phenomena. The discovery of high-$T_c$ superconductivity in these systems overturned the Bardeen-Cooper-Schrieffer (BCS) paradigm that magnetism is detrimental to Cooper pairing in conventional superconductors. Although long-range magnetic order is typically antithetical to superconductivity, magnetic or spin fluctuations are now widely recognized as a key driver of magnetically mediated pairing in unconventional superconductors [4-6]. Among 3*d* transition-metal materials, Cr-based compounds hold a special position, as superconductivity is rare due to the strong magnetic moments generally carried by Cr ions. Unsurprisingly therefore, Cr-based superconductors were not reported until 2014, when the first Cr-based compound CrAs was found to be superconducting in the vicinity of an antiferromagnetic quantum critical point driven by pressure [7]. Since then, superconductivity has been discovered in $CrSiTe_3$ [8], $CrSbSe_3$ [9], and $CsCr_3Sb_5$ [10] after the suppression of magnetic order under pressure. Notably, compounds such as $A_2Cr_3As_3$ ($A$=alkali metal) [11-13] and $Pr_3Cr_{10-x}N_{11}$ [14] exhibit superconductivity even without long-range magnetism, yet retain strong spin fluctuations and signatures of unconventional pairing [13, 15-18]. Collectively, superconductivity in Cr-based compounds is broadly regarded as unconventional, with spin fluctuations playing a prominent role in Cooper pairing. Nevertheless, the microscopic mechanisms remain elusive, particularly concerning the intricate interplay between superconductivity and magnetism.

Another remarkable advance in condensed matter physics is the concept of topological order and its material realizations, which transcends Landau's paradigm of phase transitions governed by spontaneous symmetry breaking. The interplay between topological quantum states and superconductivity can give rise to topological superconductivity with emergent Majorana fermions, the manipulation of which lays the foundation for future applications in fault-tolerant topological quantum computing [19-22]. One promising route to realize such states is to induce intrinsic superconductivity in topological materials, either through chemical doping/intercalation or via the application of pressure [19-23]. Moreover, incorporating magnetism into topological systems further enriches the physics, giving rise to magnetic topological materials [24], where intrinsic magnetism interacts with nontrivial band topology. This naturally inspires an intriguing strategy: if superconductivity can be induced in a topological material hosting magnetic order or strong spin fluctuations, the resulting system may unify unconventional and topological superconductivity within a single platform. Such a scenario is anticipated to generate novel superconducting states and exotic quantum phases, offering a unique venue to probe the synergy between unconventional and topological superconductivity. In practice, however, suitable material candidates are still scarce. A few notable examples include $FeSe_{1-x}Te_x$ [25] and $CaKFe_4As_4$ [26], unconventional Fe-based superconductors that are widely regarded as strong candidates for topological superconductivity with compelling evidence for Majorana modes, as well as Mn-based topological insulator $MnSb_4Te_7$ [27], where superconductivity emerges under pressure following the suppression of magnetic order. Yet, despite these advances, superconductivity in Cr-based materials possessing intrinsic nontrivial topology has been scarcely investigated.

The material of interest in this work is the transition-metal tetraphosphide $CrP_4$, first synthesized in 1972 [28]. Although discovered at an early stage, it has remained relatively unexplored, largely due to the experimental challenges in synthesizing phosphorus-rich pnictides, which typically require high-pressure and high-temperature (HPHT) conditions. Structurally, $CrP_4$ consists of zigzag Cr chains sandwiched between two buckled phosphorus layers, which can be viewed as a derivative of black phosphorus (BP) along the *b*-axis. Each Cr atom is coordinated by six P atoms, forming a distorted edge-sharing $CrP_6$ octahedral aligned along the *c*-axis (Fig. 1(a)). The presence of buckled BP layers suggests that $CrP_4$ may inherit certain physical characteristics of BP, which is known to undergo a pressure-driven topological quantum phase transition from a semiconductor to a topological semimetal [29, 30], followed by the emergence of superconductivity [31, 32]. The light phosphorus atoms may further play a role analogous to hydrogen in compressed hydrides [33, 34], where light elements with high Debye frequencies can significantly enhance superconductivity. Indeed, pressure-induced superconductivity has been reported in other phosphorus-rich compounds, such as $GeP_3$ [35], $GeP_5$ [36], and the structural homologues $MoP_4$ [37] and $VP_4$ [38]. Moreover, $CrP_4$ conforms to the recently proposed MHFV (Magnetic moments plus High-Frequency Vibrations) model [39], in which light elements can disrupt super-exchange magnetic interactions in compounds combining 3*d* magnetic transition-metals with light elements, thereby suppressing long-range magnetic order while mediating spin fluctuations that facilitate Cooper pairing. Notably, theoretical calculation have predicted that $CrP_4$ is a topological insulator [40], a rarity among Cr-based materials with nontrivial band topology. Taken together, the intrinsic topological character of $CrP_4$, combined with its structural features favorable for superconductivity, makes it an appealing candidate in which external stimuli such as pressure may uncover novel Cr-based topological superconductors. This prospect motivated us to investigate its high-pressure physical properties with the goal of uncovering emergent quantum phenomena and shedding light on the interplay among magnetism, superconductivity, and topology.

In this study, we synthesized $CrP_4$ single crystals using the HPHT method and conducted *in-situ* high-pressure investigations of their structural and electronic properties. Upon compression, $CrP_4$ retains its crystalline structure up to approximately 70 GPa, beyond which it undergoes an irreversible transition to an amorphous phase. Moreover, we observed a sequence of pressure-driven ground state evolutions, from metallic to semiconducting-like, and finally to a metallic state with superconductivity situated in the amorphous phase. Theoretical calculations reveal multiple topological phase transitions in the crystalline phase, initially from a strong topological insulator to a trivial state, and then back to a strong topological state under pressure. These findings identify $CrP_4$ as the first Cr-based amorphous superconductor, expanding the landscape of Cr-based superconductivity and providing a promising platform for exploring the potential interplay among superconductivity, topology, and structural disorder.

## II. EXPERIMENTAL AND COMPUTATIONAL DETAILS

$CrP_4$ single crystals were synthesized via the HPHT method using a cubic high-pressure apparatus (SPD-6×1800). High-purity chromium powder (99.99 *wt*.%) and red phosphorus pieces (99.99 *wt*.%) were employed as starting materials in a molar ratio of 1:4. The mixtures were ground to ensure homogenization, pressed into pellets, and subsequently placed into the HPHT chamber. These steps were conducted inside a glovebox under an argon atmosphere. The sample assembly, previously detailed in our earlier work [41], was heated to 1673 K at approximately 5 GPa for 3 hours, followed by rapid cooling to room temperature. Finally, single crystals of $CrP_4$, predominantly 50-200 μm in size and exhibiting a black metallic appearance, were successfully mechanically separated from the reaction product.

The crystal structure of $CrP_4$ at ambient pressure was analyzed using x-ray diffraction (XRD) with a Rigaku SmartLab SE diffractometer and Cu-$K_\alpha$ radiation ($\lambda$ = 1.5418 Å). High-resolution transmission electron microscopy (HRTEM) images were captured on a JEM2100 Plus transmission electron microscope operating at 200 kV. Elemental analysis was made by a scanning electron microscope, equipped with energy dispersive x-ray (EDX) spectroscopy probe (Zeiss, Sigma 500). The resistivity measurements were performed in a Quantum Design Physical Property Measurement System (PPMS-9T). Magnetic measurements were conducted using the vibrating sample magnetometer (VSM) module of a PPMS-9T system.

The high-pressure electrical transport measurements were conducted using a screw-pressure-type diamond anvil cell (DAC) made of non-magnetic Be-Cu alloy, equipped with 200 μm (run 1 and run 4) and 100 μm (run 2 and run 3) anvil culets. A cubic BN/epoxy mixture serves as an insulating layer between the BeCu gaskets and the electrical leads. Four Pt electrodes were attached to the sample with a four-probe van der Pauw method. A detailed schematic of the DAC assembly used for the transport measurements is shown in Fig. S1 [42]. Hall effect was performed by reversing the field direction and antisymmetrizing the data. The electrical transport measurements were performed in a PPMS-9T system. For high-pressure XRD measurements, two sets of experiments were conducted on $CrP_4$ powders in symmetric DACs. In run 5, *in-situ* laboratory-based XRD measurements were performed using a transmission wide-angle diffractometer (Rigaku, NANOPIX-WE) with Mo $K_\alpha$ radiation ($\lambda$ = 0.7093 Å) in a DAC with 300 μm anvil culets. In run 6, *in-situ* synchrotron-based XRD measurements were carried out at beamline BL15U1 of the Shanghai synchrotron radiation facility ($\lambda$ = 0.6199 Å). 150 μm anvil culets were used for run 6. The two-dimensional diffraction patterns were integrated into one-dimensional profiles using the DIOPTAS program[43]. The lattice parameters were determined by Rietveld refinements using the GSAS-II software packages [44]. High-pressure Raman experiments (run 7 and run 8) were conducted using a high-resolution Raman spectrometer (Horiba, LabRAM HR Evolution) with an excitation wavelength of 532 nm. In all high-pressure experiments, the pressure was calibrated using the ruby fluorescence shift [45] or the Raman signal of the diamond anvil [46]. Additionally, Au powders were loaded as an internal pressure standard for synchrotron XRD measurements [47].

The electronic structure of $CrP_4$ was investigated using the first-principles DFT calculations implemented within the WIEN2K package [48], employing the full-potential linearized augmented plane wave (FLAPW) method. We utilized the Wu-Cohen generalized gradient approximation (GGA) [49] for the exchange-correlation functional, which has been shown to provide superior accuracy in determining crystal structure parameters compared to conventional functionals. To ensure a comprehensive and accurate description of the electronic structure, both relativistic effects and spin-orbit coupling (SOC) were incorporated in all calculations. The Fermi surface was mapped using a high-density 24×24×19 $k$-point mesh across the complete Brillouin zone. For detailed analysis of the band topology and surface states, we constructed a tight-binding model using maximally localized Wannier functions (MLWFs) [50] as implemented in the Wannier Tools package [51]. The tight-binding Hamiltonian was built using 68 bands, incorporating Cr 3$d$ and P 3$p$ orbitals as the basis for the topological analysis. To evaluate the transport properties, we employed BOLTZTRAP [52] which implements the semi-classical approach given by the solution of Boltzmann's equation in the relaxation time approximation.

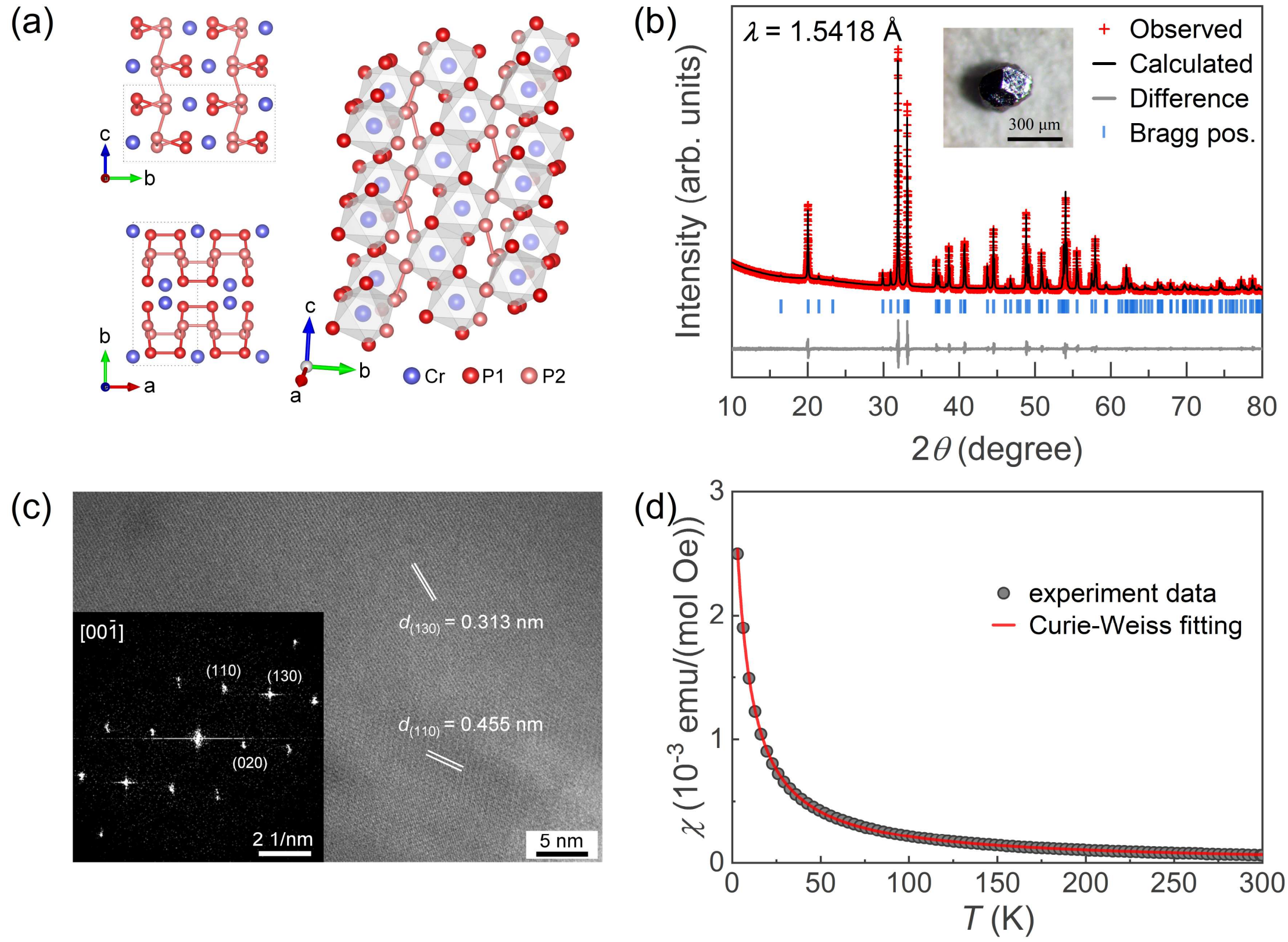

FIG. 1. Structural and physical properties of $CrP_4$ at ambient conditions. (a) Crystal structure of $CrP_4$, visualized using VESTA software [53]. (b) Powder XRD pattern obtained with Cu $K_\alpha$ radiation. The solid black line represents the Rietveld refinement. Inset: photograph of a typical $CrP_4$ single crystal. (c) HRTEM image along the $[00\bar{1}]$ direction, with the corresponding FFT pattern shown in the inset. (d) Temperature dependence of magnetic susceptibility under a 3 T magnetic field. The red line represents the Curie-Weiss law fit.

## III. RESULTS AND DISCUSSION

Figure 1(b) presents the typical XRD pattern, that all the characteristic peaks can be well indexed to a monoclinic structure with the space group *C*2/*c*, in good agreement with previous report [28]. Rietveld refinements yields the lattice parameters of $a$ = 5.19165 Å, $b$= 10.76071 Å, $c$ = 5.77036 Å, and $\beta$ = 110.6455°. EDX analysis indicates a uniform distribution of Cr and P elements in the produced crystals, with an atomic ratio of ~1:4 (19.91:80.09) (Fig. S2 [42]). Figure 1(c) illustrates the HRTEM image taken along the $[00\bar{1}]$ direction, together with the corresponding fast Fourier transform (FFT) in the inset, which confirms the crystalline nature of the produced $CrP_4$ samples. We can clearly see distinct lattice planes with inter-planar distances of 0.313 nm and 0.455 nm, corresponding to the (130) and (110) planes of the monoclinic crystal structure, respectively, consistent with the XRD results shown in Fig. 1(b). The combined results obtained from XRD, EDX, and HRTEM experiments confirm the successful synthesis of high-quality $CrP_4$ single crystals. Figure 1(d) presents the temperature dependence of magnetic susceptibility from 2 K to 300 K, which exhibits a paramagnetic ground state. By applying Curie-Weiss fit, $\chi(T) = \chi_0 + C/(T-\theta)$, we determined an effective moment of approximately 0.44 $\mu_B$/Cr. The negative value of $\theta \sim$ -6.5 K indicates the correlations between Cr ions are antiferromagnetic in essence.

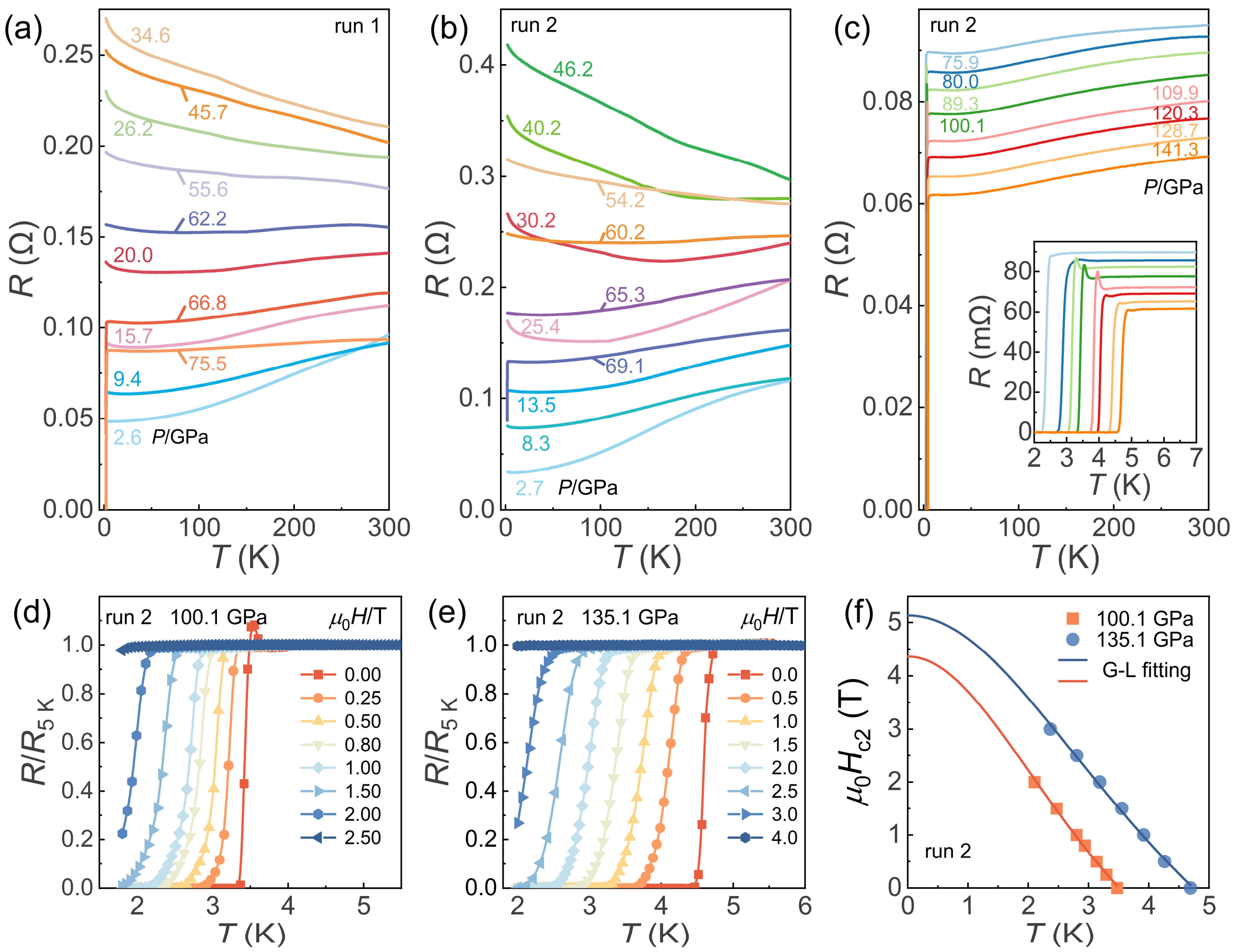

FIG. 2. Temperature-dependent resistance of $CrP_4$ under pressure. (a) Resistance evolution with pressure up to 75.5 GPa in run 1. (b) and (c) Evolution of resistance with pressure up to 141.3 GPa in run 2. Inset in (c) magnifies the superconducting transition region. (d) and (e) Resistance versus temperature under various magnetic fields at 100.1 GPa and 135.1 GPa in run 2. (f) Upper critical fields $\mu_0 H_{c2}(T)$ as a function of temperature at different pressures. $T_c$ is defined as the onset of the resistance transition, determined by the intersection of the extrapolated normal-state resistance and the steep transition curve. Solid lines represent fits to the G-L equation.

We employed diamond anvil cell (DAC) techniques to probe the transport properties of $CrP_4$ under compression. Figure 2(a) presents the temperature-dependent resistance (*R-T*) under pressures up to 75.5 GPa in run 1. At low pressures, the *R-T* curves exhibit typical metallic behavior, consistent with previous ambient-pressure observations [28, 54]. With increasing pressure, however, an unexpected transition from metallic to semiconducting-like states was observed. Specifically, in the pressure range of 15.7-20 GPa, the resistance rises obviously at lower temperatures, indicating a transition to a semiconducting-like state. This region expands significantly with further compression, and by 26.2 GPa, the metallic behavior is completely absent, leaving the semiconducting-like state across the entire temperature range. The magnitude of resistance continues to increase, reaching a maximum near 34.6 GPa. Beyond this point, both the semiconducting-like behavior and the overall resistance decrease with further compression. From 66.8 GPa, the metallic behavior in the whole temperature region reemerges as pressure increases, and meanwhile, there is a significant drop in resistance at low temperatures, eventually reaching a near-zero resistance at 75.5 GPa, signalling the onset of superconductivity.

To further examine the resistance behavior and confirm superconductivity at higher pressures, we extended the measurements to 141.3 GPa in run 2 (Figs. 2(b) and 2(c)) and 135 GPa in run 3 (Fig. S3 [42]). In run 2, the *R*-*T* curves exhibit the same pressure-dependent trend as in run 1, maintaining metallic behavior above 69.1 GPa. Notably, zero resistance was achieved at 2.2 K under 75.9 GPa, confirming superconductivity. The superconducting transition temperature $T_c$ increases steadily with pressure, reaching 4.8 K at 141.3 GPa (inset of Fig. 2(c)). Notably, a slight upturn in the *R*-*T* curve appears prior to the superconducting transition. Such behavior has been frequently observed in high-pressure resistance measurements using DAC techniques and is often attributed to a fragile precursor to superconductivity, possibly associated with pressure inhomogeneity near the superconducting transition [37, 55, 56]. In run 3, the results aligned with those obtained in runs 1 and 2, demonstrating their reproducibility. The low-temperature *R*-*T* curves under magnetic fields up to 9 T at 100.1 and 135.1 GPa (Figs. 2(d) and 2(e)) show a systematic suppression of superconductivity with increasing field. Figure 2(f) depicts the temperature dependence of the upper critical field $\mu_0H_{c2}(T)$. By fitting the data with the Ginzburg-Landau (G-L) formula, $\mu_0H_{c2}(\mathrm{T})=\mu_0H_{c2}(0)(1-(T/T_c)^2)/(1+(T/T_c)^2)$, we determined $\mu_0H_{c2}(0)$ values of 4.4 T and 5.1 T at 100.1 GPa and 135.1 GPa, respectively.

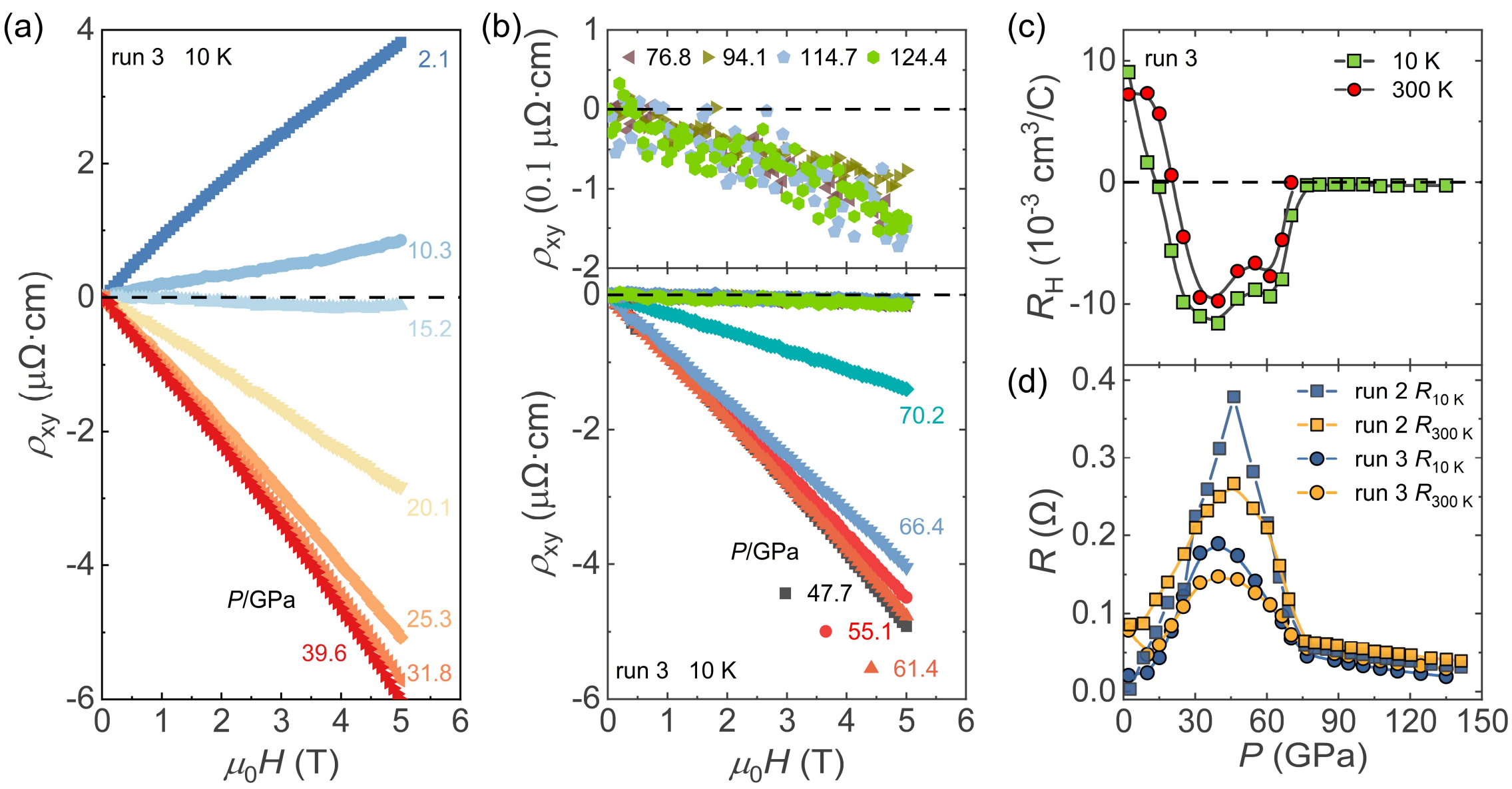

FIG. 3. Hall effect of $CrP_4$ at 10 K under pressure. (a) and (b) Plots of Hall resistivity $\rho_{xy}$ versus magnetic field $H$ under various pressures. (c) Pressure dependence of the Hall coefficient $R_H$, displaying a sign change from positive to negative at the critical pressure of around 15 GPa. (d) The resistance at 10 K and 300 K as a function of pressure for run 2 and run 3. The horizontal dashed lines in (a)-(c) serve as guides to the eye, representing the zero baseline.

The pressure-induced quantum phase transition from metallic to semiconducting-like behavior in $CrP_4$ is unusual, as external pressure generally broadens the electronic bandwidth and enhances metallicity. Similar phenomena have been reported in compressed cuprates [57] and transition-metal dichalcogenides [58, 59], albeit with distinct underlying mechanisms. To elucidate the origin of this anomalous transition, *in-situ* high-pressure Hall effect measurements were performed at 10 K (Figs. 3(a) and 3(b)) and 300 K (Fig. S4 [42]). The Hall resistivity $\rho_{xy}(H)$ displays a quasi-linear field dependence, with its slope evolving from positive to negative near 15 GPa and reaching a minimum at approximately 39.6 GPa before reversing upon further compression (Fig. 3(b)). We extracted the Hall coefficient ($R_H$) values by fitting the $\rho_{xy}(H)$ curves at low fields (Fig. 3(c)), alongside resistance at 10 K and 300 K (Fig. 3(d)) for comparison. At low pressures, the positive $R_H$ indicates hole-dominated transport, consistent with that measured at ambient pressure [54]. The sign change of $R_H$ around 15 GPa marks the transition to electron-dominated conduction and a reconstruction of Fermi surface, a result supported by our theoretical calculation shown later (Fig. 5 and Fig. S5 [42]). As pressure increases, $R_H$ decreases until 39.6 GPa, accompanied by a gradual rise in resistance and a pronounced semiconducting-like behavior. Beyond 39.6 GPa, $R_H$ then starts to increase, exhibiting a plateau between 47.7 and 66.4 GPa, where the resistance and semiconducting-like features are suppressed, signifying a gradual recovery of metallic characteristic. Above 70.2 GPa, where $CrP_4$ becomes fully amorphous (as demonstrated later), the Hall signal weakens, and $R_H$ remains nearly constant. Meanwhile, the resistance maintains metallic behavior and its decrease rate with pressure slows down. These results demonstrate that the abnormal resistance evolution under compression arises primarily from pressure-driven modifications of the electronic structure.

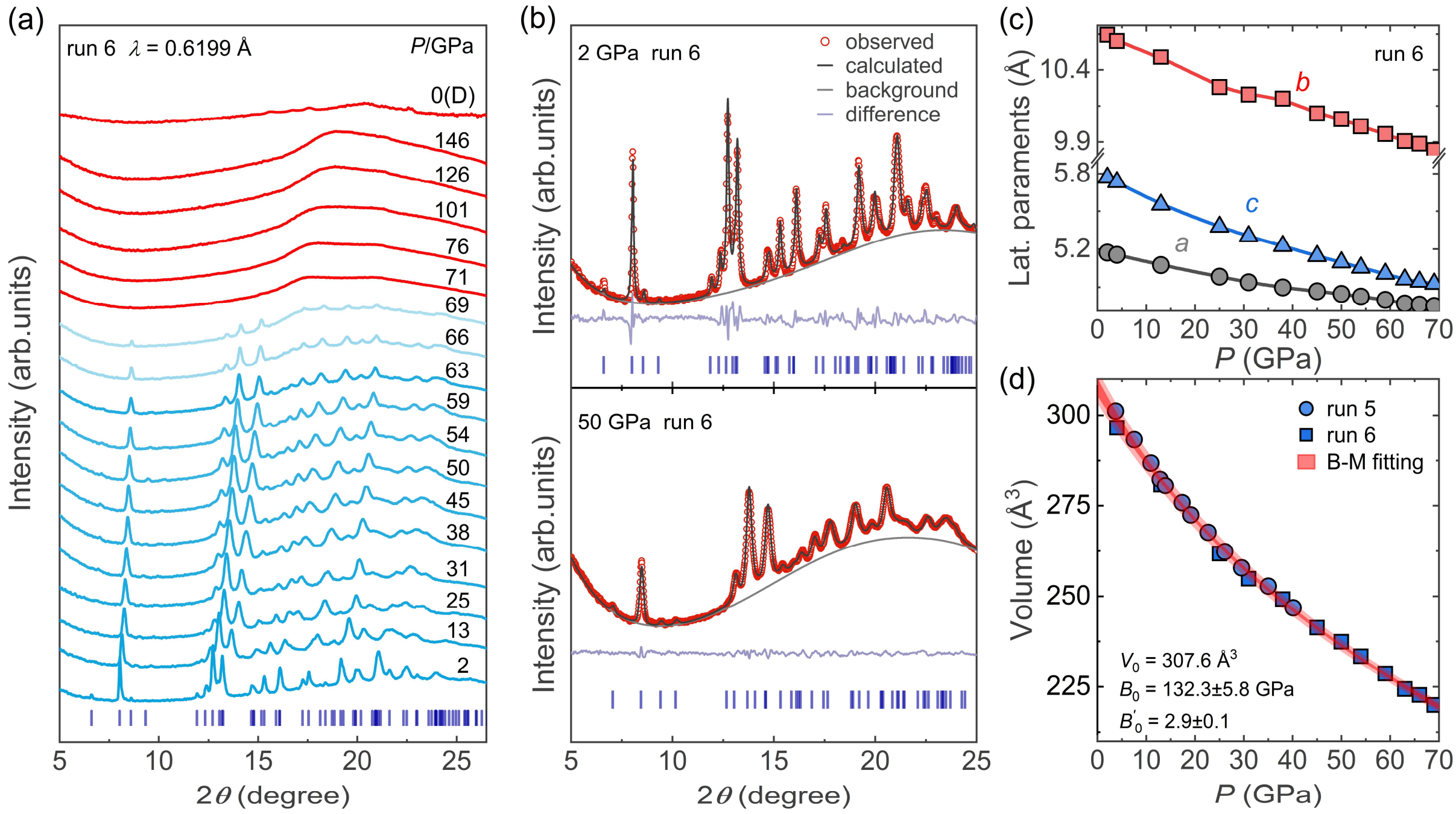

FIG. 4. *In-situ* high-pressure XRD of $CrP_4$. (a) XRD patterns at various pressures up to 146 GPa in run 6. Topmost pattern, labelled as 0(D), taken after pressure release to ambient conditions. (b**)** Representative Rietveld refinement results for XRD patterns at 2 GPa (up) and 50 GPa (bottom), respectively. (c) Refined lattice parameters ($a$, $b$, and $c$) as a function of pressure. (d) Pressure dependence of the derived cell volume $V$. Solid red line represents the fit of the experimental data to the third-order B-M equation of state.

Next, to explore the origin of pressure-induced superconductivity, we performed two *in-situ* high-pressure XRD measurements. The first utilized laboratory-based XRD (run 5) at pressures up to 43.4 GPa (Fig. S6 [42]), while the second employed synchrotron-based XRD (run 6) reaching 146 GPa. Figure 4(a) presents the XRD patterns from run 6 at various pressures, with all patterns up to 69 GPa corresponding to the pristine monoclinic structure with space group *C2/c*. Figure 4(b) illustrates representative refinements at 2 and 50 GPa, showing that as pressure increases, the diffraction peaks systematically shift to higher angles, indicating lattice contraction. Interestingly, we note that the ambient diffraction peak at approximately $2\theta \sim 16°$ begins to split above 25 GPa. This splitting does not signal a structural phase transition. Instead, it results from the different shift rates of the (220) and (13-2) peaks towards higher angles [60], which is attributed to the $c$ axis being significantly more compressible than the $a$ and $b$ axes (Fig. S7 [42]). Structural refinements reveal a gradual decrease in lattice parameters $a$, $b$, and $c$ with increasing pressure (Fig. 4(c)). The pressure-volume ($P$-$V$) data (Fig.

4(d)) fit well to the third-order Birch-Murnaghan (B-M) equation of state [61], $P(V) = \frac{3}{2}B_0\left[\left(\frac{V_0}{V}\right)^{7/3} - \left(\frac{V_0}{V}\right)^{5/3}\right] \times \left\{1 + \frac{3}{4}(B_0' - 4)\left[\left(\frac{V_0}{V}\right)^{2/3} - 1\right]\right\}$, where $V_0$ is the unit cell volume at zero pressure, $B_0$ is the bulk modulus, and $B_0'$ is the first-order derivative of the bulk modulus. The fitting results yield $V_0 = 307.6$ Å$^3$, $B_0 = 132.3 \pm 5.8$ GPa, and $B_0' = 2.9 \pm 0.1$. These findings exclude the possibility that the transition from metal to semiconducting-like states is driven by a structural transition, instead pointing to an electronic origin.

It is clearly noted that the gradual weakening of diffraction peaks above 66 GPa signifies the onset of amorphization, culminating in their disappearance near 71 GPa. This crystalline to amorphous transition well coincides with the emergence of re-metallization and superconductivity, underscoring their intimate correlation. Upon decompression, the amorphous phase persists down to the ambient pressure, as evidenced by the uppermost curve in Fig. 4(a) and the TEM measurement (Fig. S8 [42]), confirming the irreversibility of the transition. Such a pressure-induced irreversible crystalline-to-amorphous transition is further supported by *in-situ* high-pressure Raman spectroscopy measurements (Fig. S9 [42]). The quenchable amorphous phase of $CrP_4$ raises the question of whether its high-pressure superconductivity remains at ambient conditions. To address this, *R-T* curves were measured during decompression in run 2 (Fig. S10 [42]). As pressure decreased to around 63 GPa, $T_c$ decreased monotonically. However, further measurements at lower pressures were impeded by diamond culet fractures induced by rapid stress release, which damaged the Pt electrodes. To achieve a controlled pressure release, a subsequent high-pressure experiment (run 4) was performed using larger 200 μm culets. The results unambiguously show that amorphous $CrP_4$ does not retain superconductivity upon decompression (Fig. S11 [42]).

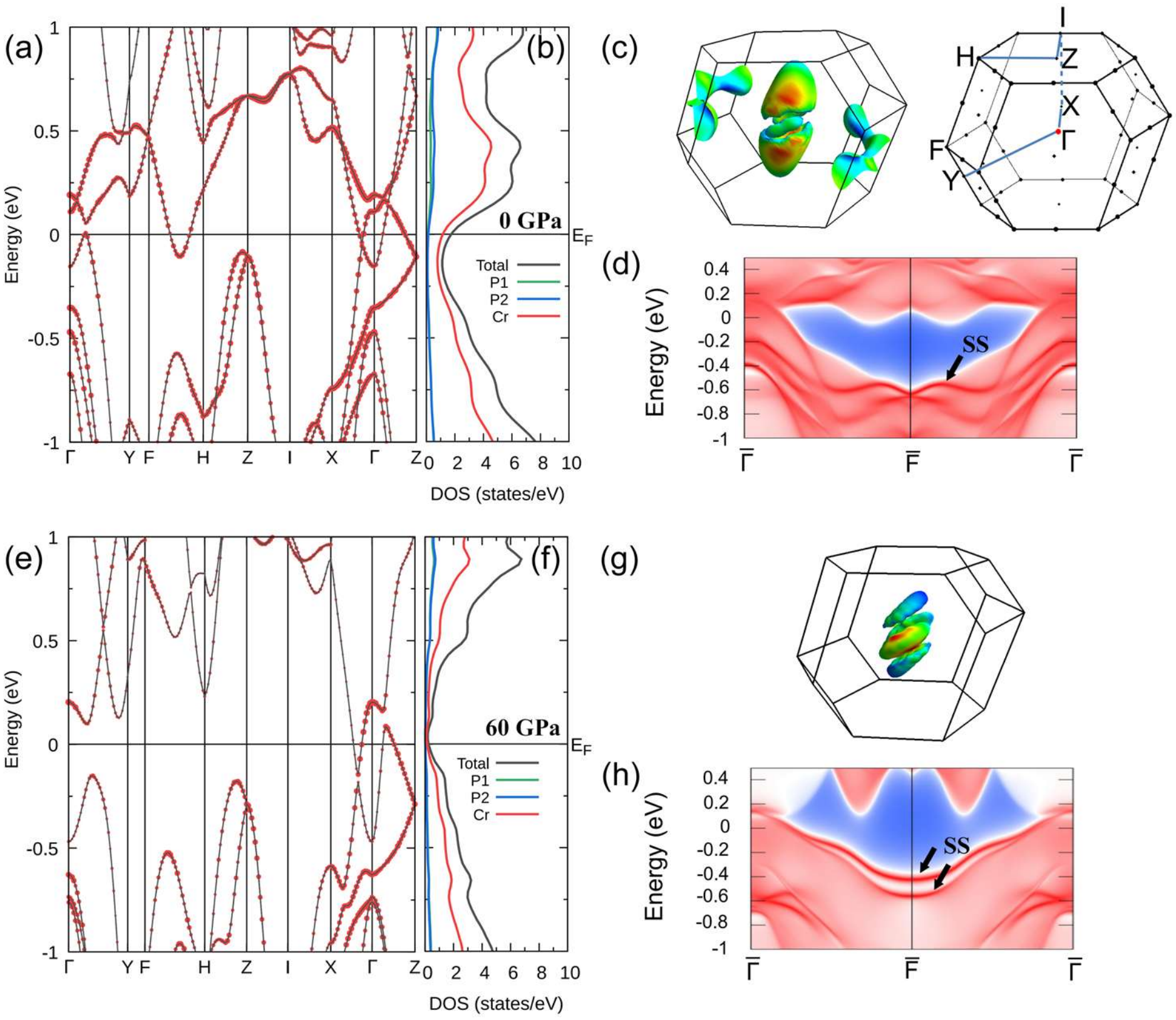

FIG. 5. Theoretical electronic properties of $CrP_4$. (a) Band structure at ambient pressure along high-symmetry path Γ-Y-F-H-Z-I-X-Γ-Z in the BZ. (b) DOS at ambient pressure, highlighting element contribution near the Fermi level, contains two inequivalent P sites (P1, P2) at Wyckoff position 8*f*

with coordinates (0.274, 0.089, 0.678) and (0.274, 0.221, 0.190). (c) Fermi surfaces and Brillouin zone. The Fermi surfaces are colored by Fermi velocity magnitude $|v_F|$ (blue: low, red: high). The high-symmetry points are marked as: Γ (0.0, 0.0, 0.0), Y (0.3082, 0.3082, -0.5456), F (0.3633, 0.3633, -0.3928), H (0.2531, 0.2531, 0.3015), Z (0.00, 0.00, 0.50), I (0.50, -0.50, 0.50), and X (0.50, -0.50, 0.00). (d) Surface states at ambient pressure along $\overline{\Gamma}$(0.0, 0.0) - $\overline{\mathrm{F}}$ (0.3633, 0.3633) - $\overline{\Gamma}$ (0.0, 0.0) on the [0 0 1] surface. The bright red curves denote the surface states. (e) Band structure of $CrP_4$ at 60 GPa. (f) DOS at 60 GPa. (g) Fermi surfaces at 60 GPa. (h) Surface states at 60 GPa, double bright surface states curves appear around the $\overline{\mathrm{F}}$ point.

To investigate the electronic structure evolution under pressure, we also performed density functional theory (DFT) calculations on the crystalline phase of $CrP_4$, with representative results presented in Fig. 5 and Figs. S12-S14 [42]. Two distinct bands cross the Fermi level, with Cr 3*d*-orbital states prominently influencing the electronic density of states (DOS) near the Fermi energy ($E_F$). This dominance implies *d*-electron correlation effects that could substantially affect the electronic properties. Additionally, the calculations demonstrate significant pressure-dependent features across various symmetry points in the Brillouin zone. At ambient pressure (Fig. 5(c)), the Fermi surface displays several distinctive features: broken-egg-shaped hole pockets split into two segments at the Brillouin zone center (Γ point), and electron sheets located along the Brillouin zone edge and between the split hole pockets at the zone center. This intricate multi-sheet Fermi surface architecture indicates the presence of multiple charge carrier types contributing to electronic transport. From ambient pressure to 20 GPa (Fig. S12 [42]). the Fermi surface undergoes significant morphological changes, indicating substantial alterations in the electronic structure. Notably, both hole and electron pockets at the Brillouin zone center experience considerable volume reduction, with the electron pocket at the Brillouin zone edge eventually disappearing. This transition suggests a pressure-induced modification in the transport properties, coinciding with the observed abnormal resistance transition and sign change in $R_H$ (Figs. 2 and 3, and Fig. S5 [42]). Intriguingly, the Fermi surface remains stable between 30 GPa and 60 GPa (Figs. S13 and S14 [42], and Fig. 5(g)), indicating a plateau in electronic structure evolution, which is also qualitatively consistent with the $R_H$ behavior observed in our experiments (Fig. 3(c)). Notable is the expansion of the indirect gap near the Fermi level along the X-Γ-Z high-symmetry direction, indicating fundamental alterations in the band structure. This gap widening may result from a reduction in the DOS at the Fermi level. Changes in the Fermi surface and band structure could serve as a precursor to pressure-induced phase transitions or novel electronic states.

Considering the topological insulator character of $CrP_4$ at ambient pressure predicted by theoretical calculations [40], we further performed a systematic theory-driven investigation of the evolution of its topological properties under pressure by calculating the $Z_2$ topological invariants ($\nu_0$; $\nu_1\nu_2\nu_3$) within the crystalline phase. These indices were computed by tracking the Wannier charge centers (WCCs) evolution across six time-reversal invariant momentum (TRIM) planes ($k_i$=0, π; $i$= $x, y, z$), offering a complete characterization of the topological nature of a given system (see Figs. S15-S19 for WCCs under different pressures [42]). The WCC method, based on the modern theory of polarization, provides a robust approach to determine the $Z_2$ invariants without requiring specific knowledge of the Wannier functions. Under ambient pressure, $CrP_4$ is identified as a strong topological insulator with $Z_2$ indices (1; 000). The non-zero $\nu_0$ signifies the presence of topologically protected surface states and an odd number of band inversions at the TRIM points. Upon compression to 30 GPa, the system undergoes a topological phase transition to a trivial insulating state with $Z_2$ indices (0; 000). Notably, further compression to 60 GPa triggers a re-entrant topological phase transition, reverting the system to a strong topological insulator phase with $Z_2$ indices (1; 000). Although the Fermi surface morphology remains largely unchanged between 30 GPa and 60 GPa, the topological properties are modified drastically, in close analogy to what was observed in $Bi_2Se_3$ and $Sb_2Se_3$ [62]. This unusual pressure-induced topological phase re-entrance suggests a complex interplay between structural parameters and electronic band topology. The reemergence of the strong topological insulator phase implies the restoration of band inversion and the associated topologically protected surface states. These pressure-induced topological phase transitions demonstrate the high tunability of the electronic structure in $CrP_4$, emphasizing the potential for pressure-tailored topological properties. The sequential transitions between topologically non-trivial and trivial states, followed by the reemergence of non-trivial topology, represent a unique phenomenon that could be exploited for pressure-controlled electronic devices.

Based on our findings, we constructed a temperature-pressure phase diagram (Fig. 6), illustrating the significant influence of pressure on the crystal structure, electronic state, and superconductivity in $CrP_4$. Under high pressure, the normal-state resistance undergoes an abnormal quantum phase transition from metallic to semiconducting-like behavior and then reverts to metallicity. Hall effect measurements and theoretical calculations indicate a major alteration in Fermi surface morphology between 15 and 20 GPa, marked by a transition in carrier type from holes to electrons. Additionally, $Z_2$ topological invariant calculations reveal a pressure-induced transition from a topologically nontrivial state at 0 GPa to trivial states at 30 GPa, and back to a nontrivial phase at 60 GPa in the crystalline phase. At approximately 70 GPa, $CrP_4$ undergoes an irreversible crystalline to amorphous phase transition, accompanied by the onset of superconductivity and reemergence of metallization. At even higher pressures, the normal-state resistance remains metallic, and superconductivity is progressively enhanced, with no saturation observed up to the maximum pressure of 141.3 GPa explored in our experiment.

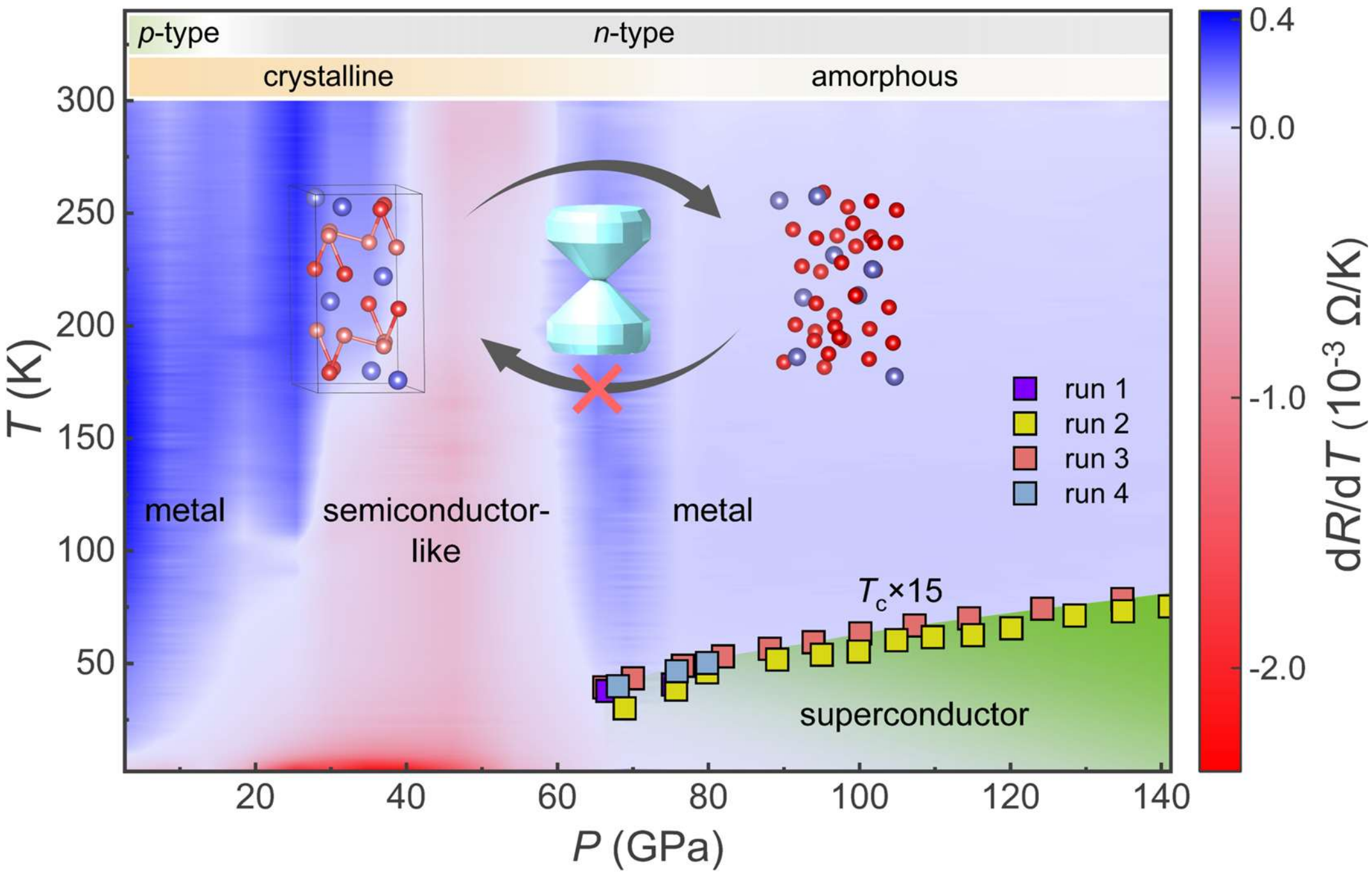

FIG. 6. Temperature-pressure phase diagram of $CrP_4$. The background mapping color illustrates the variation of $dR/dT$ with temperature under different pressures, emphasizing the sequential metal to semiconducting-like to metal transitions in resistance. The dominant carrier type changes from holes

to electrons within the 15~20 GPa range. At approximately 70 GPa, $CrP_4$ undergoes a crystalline to amorphous transition, accompanied by the onset of superconductivity. The superconducting transition temperature $T_c$, determined from the onset of the superconducting transition.

In general, Cr-containing materials exhibit long-range magnetism, with superconductivity emerging after its suppression under pressure [7-10, 63]. In such systems, spin fluctuations are widely considered essential for magnetically mediated unconventional pairing. Recently, the MHFV model was proposed to explain high-pressure superconductivity in $MnB_4$ [39]. That is, in compounds combing 3*d* magnetic-transition-metals with light elements, the latter can act as disruptors of magnetic super-exchange interactions between the magnetic ions, thereby suppressing long-range magnetic order. Under these circumstances, the local and itinerant characteristics of 3*d* electrons are delicately balanced, and the light elements mediated strong spin fluctuations are expected to play a crucial role in Cooper pairing. In $CrP_4$, the unique crystal field environment of the $CrP_6$ octahedron (Fig. 1(a)) suggests that the light phosphorus atoms may similarly disrupt the Cr-Cr super-exchange interaction, akin to the role of boron atoms in $MnB_4$ [39]. Such disruption likely accounts for the absence of long-range magnetic order in $CrP_4$. The intrinsic magnetic moments contributed by Cr ions, in turn, provide the basic source of spin fluctuations that may be crucial for Cooper-pair formation. Moreover, our theoretical calculations demonstrate that the DOS at the $E_F$ is predominantly contributed by the 3*d*-orbitals of Cr in $CrP_4$, suggesting that the superconductivity arises from the condensation of these 3*d* electrons. Thus, despite the lack of long-range magnetic order at ambient pressure, spin state transition of Cr ions (e.g., magnetic instability) and modulation of *d*-electron correlations under pressure may generate pronounced spin fluctuations, which could underlie unconventional superconductivity in $CrP_4$, analogous to that observed in paramagnetic Cr-based superconductors such as $K_2Cr_3As_3$ [11, 15, 16] and $Pr_3Cr_{10-x}N_{11}$ [14, 17]. Further investigations are needed to elucidate the pairing symmetry and microscopic mechanism underlying the amorphous superconducting state.

Furthermore, superconductivity emerges concurrently with the pressure-induced amorphization of $CrP_4$, suggesting an underlying correlation between these two phenomena. Indeed, when pressure is used to tune the ground state of a material, it often drives the system to an amorphous state where superconductivity emerges or is enhanced [64-70]. However, the mechanism by which structural disorder in amorphous systems promotes superconductivity remains poorly understood. In general, disorder tends to induce electron localization, which is seemingly incompatible with superconductivity that relies on coherent electron pairing and dissipationless charge transport. Several factors may contribute to this unusual behavior. First, recent theoretical studies suggest that disorder-induced multifractal electronic states may enhance electron correlations near the amorphous transition, potentially strengthening superconductivity in disordered systems. A similar mechanism has been proposed to account for the enhancement of superconductivity near the amorphous phase transition in $In_2Te_5$ [64]. Second, pressure-induced amorphization may strongly suppress magnetic orders. In materials containing magnetic elements such as chromium, magnetic correlations often compete with superconductivity. Structural disorder introduced during amorphization could weaken long-range magnetic interactions or magnetic ordering tendencies of Cr ions, while enhancing spin fluctuations, thereby creating a more favorable environment for superconducting pairing. Third, amorphization can substantially alter the lattice dynamics. The transition from an ordered crystalline lattice to a disordered network may alter the phonon density of states, including the appearance of additional low-energy vibrational modes and enhanced electron-phonon coupling. Such changes could strengthen pairing interactions and promote superconductivity in the amorphous phase. Taken together, the observation of superconductivity in pressure-amorphized $CrP_4$ represents a new finding in a compound containing magnetic element. To our knowledge, $CrP_4$ constitutes the first Cr-based amorphous superconductor, not only extending the family of Cr-based superconductors but also opening new avenues for exploring Cr-based superconductivity in amorphous materials.

Finally, the absence of translational symmetry in amorphous materials complicates the application of conventional momentum-space-based topological theories, since quantities such as the first Chern number and Berry curvature are generally ill-defined. As a result, experimental investigations of topological phenomena in amorphous systems remain are in its infancy [71], with only a few theoretical studies on topological insulators and semimetals [72-75]. In the present work, the topological character of $CrP_4$ is predicted from theoretical calculations for the crystalline phase under pressure, whereas superconductivity emerges only in the high-pressure amorphous phase. Based on the current experimental and theoretical results, the relationship between the topological properties identified in the crystalline phase and the superconductivity observed in the amorphous phase remains unclear. At present, we do not have direct evidence that the topological band structure of the crystalline phase leaves a measurable "legacy" effect in the amorphous superconducting state. Nevertheless, recent studies suggest that certain topological characteristics may persist in amorphous systems [71-75]. Motivated by this possibility, and considering that the amorphous phase of $CrP_4$ can be quenched to ambient pressure, this material may provide an interesting platform for exploring potential connections among disorder, topology, and superconductivity, although such a scenario remains speculative at present.

## IV. CONCLUSION

In conclusion, our comprehensive study of $CrP_4$ under pressure has revealed significant transformations in its crystal structure, electronic properties, and band topology, culminating in the discovery of the first amorphous Cr-based superconductor. The ground state undergoes quantum phase transition from a metallic phase to a semiconducting-like state, and finally to a metallic phase with superconductivity in the amorphous state. Theoretical calculations demonstrate substantial changes in the Fermi surface and the presence of multiple topological phase transitions under pressure. The discovery of superconductivity in pressurized topological insulator $CrP_4$ expands the family of Cr-based superconductors, providing a novel system for exploring the potential interplay among topology, superconductivity, and disorder in amorphous materials.

## ACKNOWLEDGMENTS

We thank Ali Bangura, Michael Smidman, and Tianping Ying for stimulating discussions and thank LiLi Zhang and Mingzhi Yuan for their kind help during the high-pressure XRD measurements at the Shanghai synchrotron radiation facility. This work was partly supported by the National Natural Science Foundation of China (Grant Nos.12574020, 12204265, 12474017, and 12274369), the Young Scientists of Taishan Scholarship (Grant No. tsqn202408168), the Natural Science Foundation of Shandong Province (Grant Nos. ZR2022QA040, ZR2023JQ001, and ZR2023QA057), the Higher Educational Youth Innovation Science and Technology Program of Shandong Province (2023KJ202), the Natural Science Foundation of Nanjing University of Posts and Telecommunications (Grant No. NY224165), and the Zhejiang Provincial Natural Science Foundation of China (Grant No. LZ25A040003).

## DATA AVAILABILITY

The data that support the findings of this article are not publicly available. The data are available from the authors upon reasonable request.

---